# Loopholes in the interpretation of experiments on the Aharonov-Bohm effect.

Herman Batelaan and Maria Becker

*Department of Physics and Astronomy, University of Nebraska—Lincoln,
208 Jorgensen Hall, Lincoln, Nebraska 68588-0299, USA,
e-mail: hbatelaan2@unl.edu*

The independence of the Aharonov-Bohm phase shift on particle velocity is one of its defining properties. The classical counterpart to this dispersionless behavior is the absence of forces along the direction of motion of the particle. A reevaluation of the experimental demonstration that forces are absent in the AB physical system is given, including previously unpublished data. It is shown that the debate on the presence or absence of forces is not settled, and an experiment searching for dispersionless forces is proposed.

Type-I Aharonov-Bohm effects [1] showcase the guiding principle of the Standard Model, local gauge invariance [2]. The Aharonov-Bohm effect is also a corner-stone phenomenon in quantum mechanics. It is thought to establish that the vector potential (or more accurately the gauge invariant loop integral of the vector potential) can cause measurable effects even when the fields (and thus the forces) are zero [3]. It is thus claimed to elevate the relevance of the vector potential from being a helpful mathematical construct to that of having direct physical reality [4]. However, Vaidman recently reconsidered this viewpoint [5]: "…the Aharonov-Bohm effect can be explained without the notion of potentials. It is explained by local action of the field of the electron on the source of the potential."  The passing electron is shown to exert a force on the solenoid, while the solenoid does not exert a force on the passing electron.  The first part of this argument agrees with the view that Boyer [6] has held. Boyer claims that there is a force on the solenoid, but in contrast, he also claims that there is a back-action force on the electron that explains the AB-phase shift. McGregor et *al.* have shown [7] that both viewpoints can be maintained even if they appear to be at odds with each other. In the case that the motion of the charge carriers in the solenoid is fully constrained, the solenoid experience a force and the passing electron does not, while if the charge carriers are completely free to move, the passing electron does experience a force. This has been shown to be an example of a Feynman paradox [7] on conservation of momentum. Missing momentum is stored in the combined electromagnetic field of the electron and solenoid in the case that there is not back-action force. When there is a back-action force, momentum conservation does not require field momentum. The surprise is that the force is exactly the correct magnitude to explain the AB-effect [5,6]. The two extreme limits, constrained and unconstrained motion, considered in ref. 7, are not thought to represent a realistic description of a physical system. However, a detailed model of the response of the solenoid is currently unknown [8]. A definitive theoretical answer is, thus, currently not available.

On the experimental side, a test showing the dispersionless nature of the Aharonov-Bohm effect with an electron wave interferometer [9,10] has never been performed. The next best approach is to rule out forces by time delay experiments. Caprez et al. have shown that an electron passing by a solenoid does not experience a force that causes a delay sufficiently large to explain the AB-effect [11,12]. It appears that this settles the issue. However, we consider two loopholes in this paper. The first considers the possibility of a different back-action for electrons in a solenoid as compared to electrons bound in atomic magnetic dipoles following ref. 6. Electrons in a conducting wire may, during the short interaction times, be effectively unconstrained, and, thus, provide a back-action [6]. But the core electrons are constrained much tighter by the atomic potential, and may, therefore, not provide a back-action. The second shows that the force being tested is approximately dispersionless and investigates the consequences thereof. Here, as well as in refs. 2 and 13, the classical concept of force and the wave concept of dispersion are combined in a semi-classical fashion. The force gives rise to a position shift, $\Delta x$, in the propagation direction of the particle. This shift can be related to a phaseshift through the expression, $\Delta \varphi = k \Delta x$, where $k$ is the wavevector. If this phaseshift is velocity independent, then the force is said to be dispersionless. Zeilinger [9] pointed out that the velocity independence of the phaseshift is a defining feature of the AB-effect, as forces would shift the position of a particle. He continued by pointing out that a dispersionless interaction does not shift the centroid position of an electron wavepacket. This view has been generally accepted [13]. But what if dispersionless forces exist? We will show that the Lorentz forces are approximately dispersionless for an electron passing a solenoid. This motivates our present reevalutation of currently proposed and performed experiments that test for the dispersionless nature or time delay.

The time delay experiment [11] is performed using a solenoid with a weak iron core. The response of the conduction electrons in the current carrying wire of the solenoid is possibly different than the iron core electrons that are bound in atomic states. Addressing the first loophole, we consider whether the experimental data of our ref. 11 is sufficient to rule out a back action that is limited to the solenoid electrons. Addressing the second loophole, we question whether the experiment is sufficient to rule out dispersionless forces.

To these ends, consider an electron passing by a current carrying solenoid. The solenoid symmetry axis is chosen to coincide with the $z$-axis, while the electron moves parallel to the $x$-axis. The $x$-component of the Lorentz force on the solenoid with cross-sectional area $A$ and magnetic field $B_0$ is given by the expression [7,14]

$$F_x = \frac{-B_0 A q v(x_e, y_e)}{4\pi} \frac{4 x_e y_e}{\left(x_e^2 + y_e^2\right)^2}, \qquad (1)$$

where $v$ is the electron velocity along the $x$-direction and $x_e$ and $y_e$ are the $xy$-coordinates of the charge relative to the solenoid's $z$-axis. Using Newton's second law this force can be integrated,

$$\Delta x = \frac{2}{m} \int_{-\infty}^{\infty} \int_{-\infty}^{t'} F_x(v, x_e, y_e) dt' dt \qquad (2)$$

to yield a relative displacement $\Delta x$ between electrons passing on opposite sides of the solenoid of $\Delta x = eB_0 A / mv_0$. Here, the approximation that $v = v_0$ is made using the assumption that the force is weak. In a semi-classical approximation the resulting phaseshift is $\Delta \varphi = k \Delta x = mv_0 \Delta x / \hbar$. It is equal to the well-known Aharonov-Bohm phase shift

$$\Delta \varphi_{AB} = \frac{e}{\hbar} \int_C \vec{A} \cdot d\vec{l} = \frac{e}{\hbar} \int \vec{B} \cdot d\vec{S}, \qquad (3)$$

which, for the case of a solenoid, gives $\Delta \varphi = eB_0 A / \hbar$. It should be emphasized that the fact that such a force can be formulated at all, is very surprising in view of the generally accepted interpretation of the effect. The proposed force does not only give rise to a phaseshift in the semi-classical approximation, but also to a time delay for electrons passing by a solenoid in the classical picture [14]. This time delay was shown experimentally not to occur in the experiment mentioned above [11].

We improve on the approximation $v = v_0$ by first calculating the effect of the force on the velocity. Combining $a_x = F_x/m$ with $a_x = \frac{dv_x}{dt} = \frac{dv_x}{dx}\frac{dx}{dt} = v_x(x_e, y_e)\frac{dv_x}{dx}$ gives $v_x = \frac{1}{m}\int F_x dx$ leading to a velocity $v^+$ ($v^-$) of the electron passing on the side with $y_e > 0$ ($y_e < 0$) of

$$v^{\pm}(x_e, y_e) = \frac{-B\alpha q}{\pi m}\int_{-\infty}^{x_e}\frac{xy_e}{\left(x^2+y_e^2\right)^2}dx$$

$$= v_0 \pm \frac{B\alpha q}{2\pi m}\frac{|y_e|}{x_e^2+y_e^2} = v_0 + \Delta v_x. \quad (4)$$

The displacement of the electron $\Delta x^{\pm} = \int \Delta v_x dt = \int(v^{\pm}-v_0)\frac{dy}{v^{\pm}}$ is given by

$$\Delta x^{\pm} = \frac{B\alpha q}{2\pi m}\int_{-\infty}^{\infty}\frac{1}{v_0\left(1+\frac{B\alpha q}{2\pi m v_0}\frac{|y_e|}{x^2+y_e^2}\right)}\frac{|y_e|}{x^2+y_e^2}dx \approx$$

$$\pm\frac{B\alpha q}{2mv_0} \mp \frac{1}{2|y_e|\pi}\left(\frac{B\alpha q}{2mv_0}\right)^2 \quad (5)$$

when $\Delta v = v^+ - v^- \ll v_0$, $\frac{B\alpha q}{2\pi m v_0}\frac{|y_e|}{x^2+y_e^2} \ll 1$, and using $\int_{-\infty}^{\infty}\frac{dx}{\left(x^2+y_e^2\right)^2} = \frac{\pi}{2y_e^3}$.

The relative displacement between electrons that pass on opposite sides of the solenoid is

$$\Delta x = \Delta x^+ - \Delta x^- = \frac{B\alpha q}{mv_0} - \frac{1}{\pi|y_e|}\left(\frac{B\alpha q}{2mv_0}\right)^2. \quad (6)$$

The semi-classical phaseshift now consists of the velocity independent AB-phaseshift and a weak velocity dependent term

$$\Delta\varphi = k\Delta x = \frac{B\alpha q}{\hbar} - \frac{1}{\hbar\pi |y_e| m v_0}\left(\frac{B\alpha q}{2}\right)^2. \qquad (7)$$

The velocity independent term (first term) would now explain the usual observed AB-phaseshift, while the second term causes the envelop of a wave packet to shift by the small amount

$$\Delta x_{semi} = \frac{\partial \varphi}{\partial k} = \frac{1}{\hbar^2 k^2 \pi |y_e|}\left(\frac{B\alpha q}{2}\right)^2. \qquad (8)$$

The relation between the magnetic field and the solenoidal current is given by

$$B = \mu_r \mu_0 n I \ , \qquad (10)$$

where $\mu_r$ is the relative permeability, $\mu_0$ is the vacuum permeability, $n$ is the number of windings per unit length, and $I$ is the current. For the case that the iron core is taken into account, the magnetic field is enhanced by approximately a factor of $\mu_r \approx 150$ [11]. For the case that the back-action of the iron core is absent, the relative permeability is set equal to one.

The classical time delay follows from the first term of Eq. (6),

$$\Delta t_{clas} = \frac{\Delta x}{v_0} = \frac{B\alpha q}{m v_0^2}, \qquad (11)$$

where the magnetic field is given by Eq. (10). The semi-classical delay follows from Eq.(8),

$$\Delta t_{semi} = \frac{\Delta x_{semi}}{v_0} = \frac{1}{\hbar^2 k^2 \pi |y_e| v_0}\left(\frac{B\alpha q}{2}\right)^2. \qquad (12)$$

In the time-delay experiment [11] an electron passed by a macroscopic solenoid. To start this time-of-flight experiment, a femtosecond laser pulse was used to extract electrons from a field emission tip [15,16]. The electron pulse then passed between two identical solenoids. The two solenoids were connected through high permeability magnet iron bars to form a square magnetic toroid. This arrangement reduces magnetic flux leakage and enhances the magnetic flux by $\mu_r$. Finally, the arrival of the electron was detected with a channelplate, and a time-of-flight spectrum was obtained.

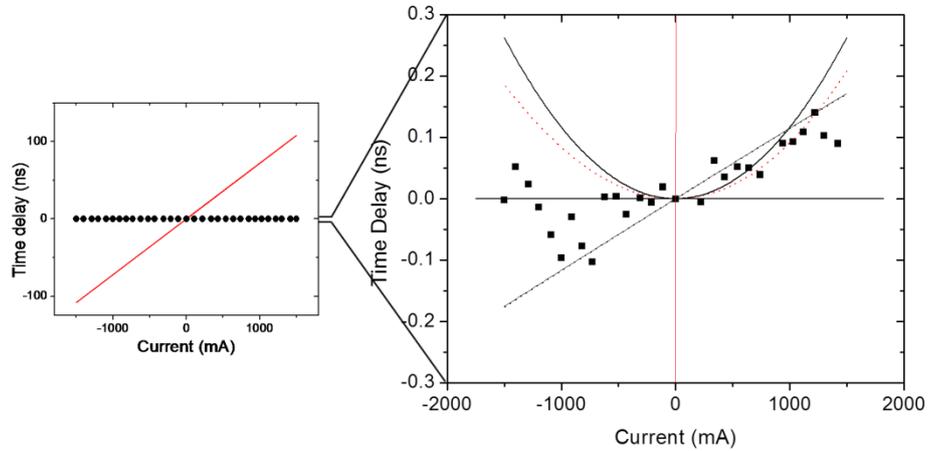

Figure 1. Time of flight data. The left panel indicates that electrons passing by a current carrying solenoid experience a time delay (black dots) that is much smaller than the predicted classical time delay (Eq. (11)) as indicated with the red solid line. The right panel shows the same data with an expansion of the time scale. The horizontal black line is the generally accepted prediction, the blue dotted line is the classical prediction without the iron core ($\mu_r = 1$), while the curved lines represent the analytic result (Eq. (12), solid line) and the numerical result (dashed line) of the semi-classical theory. The experimental data is not good enough to rule out any of the predictions.

Time a flight spectra were fitted to find the electron's arrival time. In the left panel of fig. 1 the result of ref. [11] is repeated. The experimental time delay data is compared with the classical theory (Eq. 11). It shows that no delay occurs ruling out the classical prediction. The right panel of Fig. 1 shows the same data but with the time scales expanded by three orders of magnitude. A comparison is made with the semi-classical theory without the iron core. For the applied current $I$, the 2.5 mm diameter solenoid gives a magnetic flux of $B_0 A = \mu_r \mu_0 I n A$, where $\mu_r \sim 150$ is the relative magnetic permeability of the iron core, $\mu_0$ the permeability of free space, $n = 3$/mm is the winding density, and $A = \pi r^2$ with $r = 1.25$ mm. For these parameters the classical time delay is indicated in the left panel by the solid line, while for the right panel $\mu_r = 1$. The theoretical curves are surprisingly close to the data when the iron core does not contribute. An experiment that is similar to that of ref. 11, but with improved sensitivity (about 10 times) and without an iron core, is thus proposed to rule out the classical and semi-classical theory.

It should be noted that the first experiment confirming the AB-effect, performed by Chambers [17], uses a magnetic whisker made of an iron core enclosed by the arms of an electron interferometer. If iron cores had no back-action, as considered in this paper, then the Chambers' experiment would apparently not have shown an AB effect. However, as pointed out in Chambers' paper (attributed to Pryce), the field leakage from the magnetic whisker is exactly right to explain the observed effect in terms of a classical Lorentz force. Additionally, a back-action must be absent for the leakage field explanation to hold. This also implies that an experiment with a straight, (non-tapered) magnetized iron core where leakage fields are controlled, together with the result in [11], can rule out the force explanation.

In the Möllenstedt experiment [18], electrons were passed by a small solenoid; no iron core was used. In this experiment, the back-action as proposed by Boyer, could explain the observed AB-effect. In Tonomura's famous experiment [19], the situation was more complicated. Magnetized toroids embedded in a super conducting field were used, and the AB-effect was observed. The Meisner effect was used to ensure that no magnetic leakage fields from the toroid could play a role. However, no model has been made of the response of the toroidal system to a passing electron and its potential back-action. Note that even though the Meisner effect shields the DC magnetic flux of the toroid, its shielding does not extend to fast pulsed fields (above the inverse plasmon frequency) as induced by the passing electron [20].

The crucial test of the dispersionless nature of the AB-effect has never been performed. A similar test to the one discussed below has been proposed [21] The requirement for the test is that the induced AB-phaseshift, $\varphi_{AB}$, has to exceed the coherence length (in units of $2\pi/\lambda_{dB}$): $\varphi_{AB} > 2\pi L_{coh}/\lambda_{dB}$. Because the coherence length for previous experiments was typically $10^5$ deBroglie wavelengths (Table 1), and the induced phaseshift was limited to several hundred times $2\pi$, this requirement was never met. A comparison of parameters of several experiments and a proposed experiment that meets the above requirement is given in table 1. The proposed experiment is a modification of the Mollenstedt experiment with adjusted experimental parameters. The energy is lowered to 1 keV to decrease the coherence length, which is given by $L_{coh} = \frac{\lambda^2}{\Delta\lambda} = \frac{h}{\Delta E}\sqrt{\frac{2E}{m}}$ [22] . The magnetic flux is that of a 50 micron diameter solenoid, with 12 micron diameter gold wire that supports 0.1 A current (maximum current 0.3 A). The electron interferometer with the largest beam separation ever achieved is 100 micron, which can enclose such a solenoid. Thus, the experiment is within reach of current technology.

Typically, two possible outcomes of the experiment testing the dispersionless nature are considered. A) There is no back-action in the AB-effect, and its usual interpretation is correct. In this case fringes will be observed outside the electron's coherence length. B) There is a back-action for solenoids of this type, the experiment is not an AB-effect, and fringes will not be observed outside the electron coherence length. The proposed test is generally expected to give outcome A) and demonstrate the dispersionless nature of the magnetic AB-effect [9, 23, 13, 10]. But, what if dispersionless forces exist? In this case there is a third option C). If the time delay $dt_{clas}$ in figure 1 has a value giving $vdt_{clas} > L_{coh}$, but at a lower current where $dt_{semi} < dt_{clas}$, then the possibility exist that $vdt_{semi} < L_{coh}$. In this case the observation of fringes rules out classical forces, but not the existence of semi classical forces. For an experiment to rule out dispersionless forces the current must be high enough that $vdt_{semi} > L_{coh}$. For all previous experiments (see table 1) using Eq. 11 leads to $vdt_{clas} < L_{coh}$. For example, Tonomura's famous experiment has $vdt_{clas} \approx 2 \times 10^{-11} m << L_{coh} \approx 3 \times 10^{-6} m$. The proposed experiment has $vdt_{clas} \approx 10^{-6} m > L_{coh} \approx 10^{-8} m$, but using Eq. 12 gives $vdt_{semi} \approx 10^{-9} m < L_{coh} \approx 10^{-8} m$. To rule out dispersionless forces interference experiments need to be pushed to even higher enclosed magnetic fluxes.

| Experiments | Electron Energy (keV) | deBroglie Wavelength (pm) | Coherence Length (nm) | Phase Shift (π×radians) | Shift (nm) | Magnetic Flux (G*cm$^2$) |
|---|---|---|---|---|---|---|
| Chambers [17] | 20 | 8.7 | 1200 | 800 | 3.5 | 1.7×10$^{-4}$ |
| Mollenstedt [18] | 40 | 6.1 | 1632 | 2 | 0.0061 | 4.1×10$^{-7}$ |
| Bayh [24] | 40 | 6.1 | 1632 | 2 | 0.0061 | 4.1×10$^{-7}$ |
| Schaal [25] | 50 | 5.5 | 1825 | 40 | 0.11 | 4.1×10$^{-7}$ |
| Tonomura [19] | 150 | 3.2 | 3200 | 5.5 | 0.0088 | 2.4×10$^{-6}$ |
| Proposed Experiment | 1 | 39 | 77 | 48000 | 940 | 9.9×10$^{-3}$ |

Table 1. Comparison of experiments with our proposed experiment.

Although the original Aharonov-Bohm effect has not been tested for its dispersionless nature, in a tour-de-force experiment of the scalar analogue of the AB-effect [26], it has been shown to be dispersionless. Does this rule out the existence of dispersionless forces? In ref. 26, it was pointed out that these results cannot be generalized to the original electron-solenoid case. Moreover, for the same question can be asked as stated above. Is there an approximately dispersionless force that could be responsible for these effects? This would require a detailed microscopic description of the interaction between both interacting constituents of the AB-effects for each case to predict the magnitude and thus test for it. Such detailed descriptions are not available in the literature, and the question whether or not dispersion forces exist can currently not be answered based on these experimental results.

In the broader context of modern field theories, it may appear that searching for forces in the AB-effect is a philosophical throwback to classical physics. After all, local gauge invariance of potentials has become a central means by which to find the interactions between particles [27]. Or, in other words, it is the "unobservability of potentials" (as they are affected by the choice of gauge field) that has become a guiding principle in particle physics [28]. On the other hand, the AB-effect is a rare, if not the only, experimental example, where a measured phenomenon depends on the loop integral of potentials for a case where the fields are zero. The loop integral is gauge invariant and thus measurable. It is no surprise that the AB-effect is mentioned in the context of field theory [29]. Given this unique position, it is important to verify that the experiments are performed correctly and no loopholes remain.

In summary, two loopholes in the interpretation of experiments on the Aharonov-Bohm effect are discussed. The first is the possibility that magnetized iron cores do not provide a classical back-action reducing the predicted time-delay. The second is the possibility that dispersionless forces exist. Both of these possibilities make the time-delay experiment inconclusive. An experiment without an iron core is proposed to rule out classical and dispersionless forces. Alternatively, an electron interferometer experiment with a non-tapered magnetized iron whisker could be performed. Finally, proposed tests of the dispersionless nature of the AB-effect can be performed in two regimes. In the first regime classical forces, and in the second, and harder to reach, regime, dispersionless forces can be ruled out.

**Acknowledgment**

This material is based upon work supported by the National Science Foundation under Grant No. 2505210148001.

**References**

[1] Aharonov, Y. and Bohm, D., Significance of Electromagnetic Potentials in the Quantum Theory. Phys. Rev. **115**, 485 (1959).

[2] Quigg, C., *Gauge Theories of the Strong, Weak, and Electromagnetic Interactions*. Frontiers in Physics, Lect. Notes Series 56 (Addison Wesley, Reading, 1983) p 37-50.

[3] Olariu, S. and Popescu, I.I., The quantum effects of electromagnetic fluxes. Rev. Mod. Phys. **57**, 339-436 (1985).

[4] Gasiorowicz, Quantum Physic, 3$^{rd}$ Ed., (John Wiley and Sons, New York, 2003), Supplements Ch. 16.

[5] Vaidman, L., Role of potentials in the Aharonov-Bohm effect. Phys. Rev. A **86**, 040101(R) (2012).

[6] Boyer, T. H., Semiclassical explanation of the Matteucci-Pozzi and Aharonov-Bohm phase shifts, Foundations of Physics **32**, 41-50 (2002).

[7] McGregor, S., Hotovy, R., Caprez, A., and Batelaan, H., On the relation between the Feynman paradox and the Aharonov–Bohm effects, *New J. Phys.* **14**, 093020 (2012).

[8] Boyer, T. H., Classical interaction of a magnet and a point charge: The Shockley-James paradox. Phys. Rev. E 91, 013201 (2015)

[9] Zeilinger A., Horne M.A., Aharonov-Bohm with Neutrons., Physics World **2**, 23 (1989).

[10] Matteucci, G., Iencinella, D., Beeli, C., The Aharonov-Bohm phase shift and Boyer's critical considerations: new experimental result but still an open subject? Found. of Phys. **33**, 577-90 (2003).

[11] Caprez, A., Barwick, B., and Batelaan, H., Macroscopic test of the Aharonov-Bohm effect, Phys. Rev. Lett. **99**, 210401, (2007).

[12] Batelaan, H. and Tonomura, A.,The Aharonov–Bohm effects: Variations on a subtle theme,  Phys. Today **62**(9), 38-43, (September 2009).

[13] Boyer, T.H., Darwin-Lagrangian analysis for the interaction of a point charge and a magnet:  considerations related to the controversy regarding the Aharonov-Bohm and Aharonov-Casher phase shifts. J. of Appl.  Phys. A:  Math. Gen. **39**, 3455-3477 (2006).

[14] Boyer, T.H., Semiclassical Explanation of the Matteucci-Pozzi and Aharonov-Bohm Phase Shifts.  Foundations of Physics, **32**,41-50 (2002).

[15] P. Hommelhoff, Y. Sortais, A. Aghajani-Talesh, and M. A. Kasevich, Field Emission Tip as a Nanometer Source of Free Electron Femtosecond Pulses. Phys. Rev. Lett. **96**, 077401 (2006)

[16]  B Barwick, C Corder, J Strohaber, N Chandler-Smith, C Uiterwaal and H Batelaan, Laser-induced ultrafast electron emission from a field emission tip. New J. Phys. **9**, 142 (2007).

[17] Chambers R G 1960 Shift of an electron interference pattern by enclosed magnetic flux. Phys. Rev. Lett. **5** 3–5

[18] Möllenstedt, G., and Bayh, W.,  Messung der kontinuierlichen Phasenschiebung von Elektronenwellen im kraftfeldfreien Raum durch das magnetische Vektorpotential einer Luftspule Naturwissenschaften **49,** 81(1962).

[19] A. Tonomura, N. Osakabe, T. Matsuda, T. Kawasaki, J. Endo, S. Yano, and H. Yamada, Evidence for Aharonov-Bohm effect with magnetic field completely shielded from electron wave, Phys. Rev. Lett. **56**, 792-795, (1986).

[20] Tinkham, Michael (1996). *Introduction to Superconductivity*. Dover Publications.

[21] S. Lepoutre, A. Gauguet, G. Trénec, M. Büchner, and J. Vigué, He-McKellar-Wilkens Topological Phase in Atom Interferometry, Phys. Rev. Lett. **109**, 120404 (2012).

[22] Tonomura, A., *Electron Holography*, (Springer, New York, 1999), p.16-17.

[23] A. Zeilinger, Generalized Aharanov-Bohm Experiments with Neutrons., Fundamental  Aspects of Quantum Theory, Como 1985, V.Gorrini, A.Figuereido (Eds.), (Plenum Press, New York, 1986), p. 311.

[24] Bayh, W., Z. Phys., 169, 492 (1962).

[25] Schaal, V.G., C. Jönsson, and E.F. Krimmel, Weitgetrennte kohärente Elektronen-Wellenzüge und Messung des Magnetflusses. Optik, **24**, 529-538, (1966/1967).

[26] G. Badurek, H. Weinfurter, R. Gähler, A. Kollmar, S. Wehinger and A. Zeilinger Nondispersive Phase of the Aharonov-Bohm Effect., Phys.Rev.Lett. 71, 307-311 (1993).

[27] Chaichian, M., Nelipa, N.F., *Introduction to Gauge Field Theories* (Springer, New York, 1984) p. 26-35.

[28] http://www.hep.princeton.edu/~mcdonald/examples/aharonov.pdf

[29] Quigg, C., Gauge Theories of the Strong, Weak, and Electromagnetic Interactions, Frontiers in Physics Lecture Series 56 (Addison-Wesley, Reading, 1983) p.43.